\documentclass[lettersize,journal]{IEEEtran}

\ifCLASSOPTIONcompsoc
  \usepackage[nocompress]{cite}
\else
  \usepackage{cite}
\fi

\ifCLASSINFOpdf
\else
\fi

\usepackage{graphicx}
\usepackage{subcaption}
\usepackage{xcolor}
\usepackage{mathtools}
\usepackage{hyperref}
\usepackage{makecell}
\usepackage{tikz}

\def\checkmark{\tikz\fill[scale=0.4](0,.35) -- (.25,0) -- (1,.7) -- (.25,.15) -- cycle;}

\begin{document}

\title{O-RAN: Analysis of Latency-critical Interfaces and Overview of Time Sensitive Networking Solutions}

\author{Esteban~Municio,
		Gines Garcia-Aviles, 
		Andres Garcia-Saavedra
        and~Xavier~Costa-Pérez 
\IEEEcompsocitemizethanks{\IEEEcompsocthanksitem Esteban Municio and Gines Garcia-Aviles are with i2CAT Foundation.
\IEEEcompsocthanksitem Andres Garcia-Saavedra is with NEC Laboratories Europe.
\IEEEcompsocthanksitem Xavier~Costa-Perez is with i2CAT Foundation, NEC Labs Europe and ICREA.}
}

\markboth{Journal of \LaTeX\ Class Files,~Vol.~14, No.~8, Oct~2023}%
{Shell \MakeLowercase{\textit{et al.}}: Bare Demo of IEEEtran.cls for Computer Society Journals}

\IEEEtitleabstractindextext{
\begin{abstract}

5G and B5G/6G foundations heavily rely on virtualization technologies, and virtualized Radio Access Networks (vRANs) are one of their major keystones. However, while vRANs have been traditionally suffering from significant hardware/software coupling, next generation vRANs aim for open, standardized interfaces and multi-vendor, interoperable components to enable truly flexible deployments following the cloud-native principles. In this line, the O-RAN Alliance is promoting a novel Open RAN architecture to further boost flexibility and cost efficiency.
In order to reduce costs and effectively achieve the promised disaggregation levels, O-RAN must ensure shared, integrated transport networks in opposition to dedicated, overprovisioned links from traditional approaches. However, keeping deterministic performance requirements in such cost-effective networks (i.e., general-purpose Ethernet networks), especially in those interfaces that are time-critical, is a challenge.
 In this article, we review the most relevant Time Sensitive Networking (TSN) standards that may bring compelling benefits to O-RAN, (i.e., IEEE 802.1CM, IEEE 802.1Qbu and IEEE 802.1Qbv) for providing determinism over cost-efficient networks.
We explore the design space for a TSN-enabled O-RAN architecture, reporting on the requirements and deployment options and finally, we discuss on the opportunities and challenges that O-RAN will face when adopting TSN technologies to fully open the vRAN ecosystem.

\end{abstract}

}

\maketitle

\IEEEdisplaynontitleabstractindextext

%
\IEEEpeerreviewmaketitle

\section{Introduction}\label{sec:introduction}
Future cellular networks are pushing towards flexible, disaggregated and open architectures to truly materialize the concepts of network softwarization under the promise of higher efficiency and reduced costs. 
New functionalities include multi-vendor network function split, on-demand network slicing provision, and real-time, end-to-end control of the physical infrastructures~\cite{bonati2021intelligence}. A major world-wide, carrier-led efforts to standardize next generation virtualized Radio Access Network (vRAN) architectures is done by the O-RAN Alliance, a consortium of industry and academia that aims for opening the rigid hardware-software coupling~\cite{garcia2021ran}.

One of the main features proposed by the new O-RAN architecture is the possibility of a fully-fledged virtualization of the network infrastructure. All components of the O-RAN architecture may be deployed in the O-RAN Cloud (O-Cloud) platform, which allows network operators to orchestrate the physical and virtual functions over a flexible cloud computing platform. 
Such O-RAN functionalities (e.g., standardized interfaces, infrastructure sharing or automated instantiation) aim to considerably reduce the operational and fixed costs for network operators. Among these, avoiding network overprovisioning and enabling cost-effective networks is fundamental to improve the efficiency of the deployments.

These are however ideas not devoid of challenges. The O-RAN \mbox{"7-2x"} functional split imposes strict latency and throughput requirements, specially in the interfaces carrying fronthaul traffic.
 While legacy and current interfaces, for example, Common Public Radio Interface (CPRI) and enhanced CPRI (eCPRI), heavily rely on such network overdimensioning at the design phase to cope with a relatively predictable traffic demand, the new sharing and openness levels of \mbox{O-RAN} may significantly modify traditional assumptions on the transport network. Therefore, advanced traffic mechanisms to provide real, strict deterministic performance over multiple time-critical flows in cost-efficient networks such as multi-purpose Ethernet networks are required. 

To provide such determinism, guaranteeing upper bounds on end-to-end latency and latency variation, Time Sensitive Network (TSN) mechanisms have been 
used in IEEE 802.1 networks since more than a decade ago~\cite{finn2018introduction}. Some TSN features such as Frame Preemption (IEEE 802.1Qbu~\cite{IEEE8021Q}) are nowadays proposed to provide strict packet prioritization in the fronthaul. However, in O-RAN use cases where transport networks are expected to be extensively shared, strict priority mechanisms are not sufficient. This is the case, for example, when high traffic load causes a significant increase in "\textit{fan-in}" delay (i.e., time-critical packets with the same priority arrive to a bridge at different ingress ports but depart from the same egress port), which may cause that the end-to-end packet delay exceeds the maximum allowed (typically 100$\mu$s) by the O-RAN OpenFronthaul~\cite{oranOFCUS}.
In such cases, advanced TSN scheduling mechanisms, such as Time-Aware Shaping (IEEE 802.1Qbv~\cite{IEEE8021Q}) may be used to ensure that multiple high-priority flows can efficiently share an Ethernet-based transport network without losing any determinism\footnote{802.1Qbv and 802.1Qbu amendments are now included in 802.1Q-2018}.

TSN will play a crucial role in reducing deployment costs in O-RAN deployments. First, the use of expensive, dedicated point-to-point fiber links in the fronthaul can be avoided by deploying a programmable, multi-purpose and inexpensive TSN-enabled Ethernet bridged network. Secondly, worldwide existing legacy Ethernet-based cellular infrastructure can be integrated within the O-RAN ecosystem by making it TSN-capable.
Indeed, the benefits of using TSN to support the fronthaul have been already acknowledged by the industry, triggering the standardization of IEEE 802.1CM for the use of TSN in the fronthaul~\cite{IEEE8021CM2018}.
Besides reducing costs, the adoption of TSN in the fronthaul enables the fronthaul-backhaul convergence (\textit{Crosshaul}), while enabling statistical multiplexing gain and avoiding the wavelength-dependence and inefficiency of traditional CPRI fronthauls~\cite{gonzalez2019integrating}.

In this paper, we revisit the well-established ideas of TSN to apply them in the context of O-RAN, seeking for opportunities and challenges that TSN standards can bring to the O-RAN ecosystem to reduce costs. We do this by first, mapping the TSN mechanisms to the O-RAN architecture, analyzing which interfaces can benefit from TSN, and 
 presenting the different integration alternatives.
Secondly, we summarize the O-RAN time-sensitive requirements and challenges, introducing then different deployment options and analyzing the alternatives for integrating TSN in the available O-RAN open-source components.
 To conclude, a discussion on the opportunities and research gaps is provided, summarizing the findings and the challenges ahead.

\section{Mapping TSN onto O-RAN}\label{sec:architectures}

\subsection{A Time-Sensitive Networking Primer}

The aim of TSN is to enable bounded latency and Packet Delay Variation (PDV), and ultra-reliable packet delivery to different flows coexisting in 802.1 Ethernet networks (low-cost and ubiquitous)~\cite{simon2018design}. 
This means that in TSN-enabled networks, the minimum latency may not necessarily be lower than in priority-based ones. However, the PDV is deterministic and limited to a certain time-window. 

Among the different existing TSN standard amendments included in the IEEE 802.1Q-2018 revision~\cite{IEEE8021Q}, there are two main TSN extensions that stand out when affecting the performance of time-critical transport: Frame Preemption (802.1Qbu) and Time-Aware Shaping (802.1Qbv).

\subsubsection{IEEE 802.1Qbu Frame Preemption}

The 802.1Qbu amendment is usually coupled with the IEEE 802.3br Interspersing Express Traffic amendment to enable the interruption of an on-going frame transmission to transmit one or more higher-priority frames. To this end, the standard  allows time-critical "preempting" frames with high priority to suspend the transmission of possibly on-going lower-priority "preemptable" frames in order to reduce the queuing delay of the preempting frame. This necessarily causes the fragmentation of the preempted frame and requires additional reassembling, but it significantly reduces the degree of interference between flows. However, there are limits to frame preemption. It occurs only if at least 60 bytes of the preemptable frame have been transmitted and at least 64 bytes (including CRC) remain to be transmitted. The worst-case preemption latency is defined by the minimum fragment size (124 bytes), which cannot be preempted~\cite{IEEE8021Q}.

\subsubsection{IEEE 802.1Qbv Time-Aware Shaping}

The 802.1Qbv queuing algorithm controls all the queues of a given port in the bridge by opening and closing the "transmission gates" following a rotating schedule. This schedule is synchronized with the other bridges' schedules and consists of an 8-bit tuple acting as a gate control list for the different ports. Traffic classes are mapped to the different queues, and only one traffic class is allowed to transmit at a given time. In order to prevent that large frames from one queue transmit during the next schedule entry, 802.1Qbv envisages the use of Guard Intervals (GI) between entries~\cite{IEEE8021Q}. This results in a TDMA-like behavior that eliminates interference between queues and highly reduces or completely removes PDV~\cite{gomes2018boosting}.

\subsection{Open RAN Architecture}

O-RAN architecture organizes LTE and NR RANs following the view depicted in Fig.~\ref{fig:arch}. It introduces two novel components, the non-Real-Time (non-RT) RAN Intelligent Controller (RIC) and the near-Real-Time (near-RT) RIC. The non-RT RIC is typically hosted by the Service Management and Orchestration (SMO) and performs management and control operations at large time scales (seconds or minutes).
On the other hand, the near-RT RIC performs management and control of the RAN at smaller time scales (between 10ms and 1s) through the \textit{xApps}, microservices supporting custom logic to optimize the management of radio resources. The near-RT RIC controls the E2 nodes, namely, O-RAN Centralized Units (O-CUs), O-RAN Distributed Units (O-DUs) and O-RAN-compliant LTE eNBs, and can be hosted at the edge co-located with the E2 nodes or fully decoupled if the latency constraints are fulfilled. This allows for different deployment flavors that can be changed over time according to the particular operator needs at any given time~\cite{garcia2021ran}.

The functional split selected by O-RAN is "7.2x", which balances O-RUs simplicity (lighter and less power-hungry) and traffic reduction in the fronthaul~\cite{garcia2018wizhaul}. This split keeps the Fast Fourier Transform (FFT), cyclic prefix and precoding (optional) in the O-RU, and the remaining functions are executed in the O-DU (MAC and RLC). O-RUs offloading the precoding to the O-DU are called Category~A and those that perform it, are Category~B.
 
\subsection{When TSN meets O-RAN}

\begin{figure}
        \centering
        \includegraphics[width=0.49\textwidth]{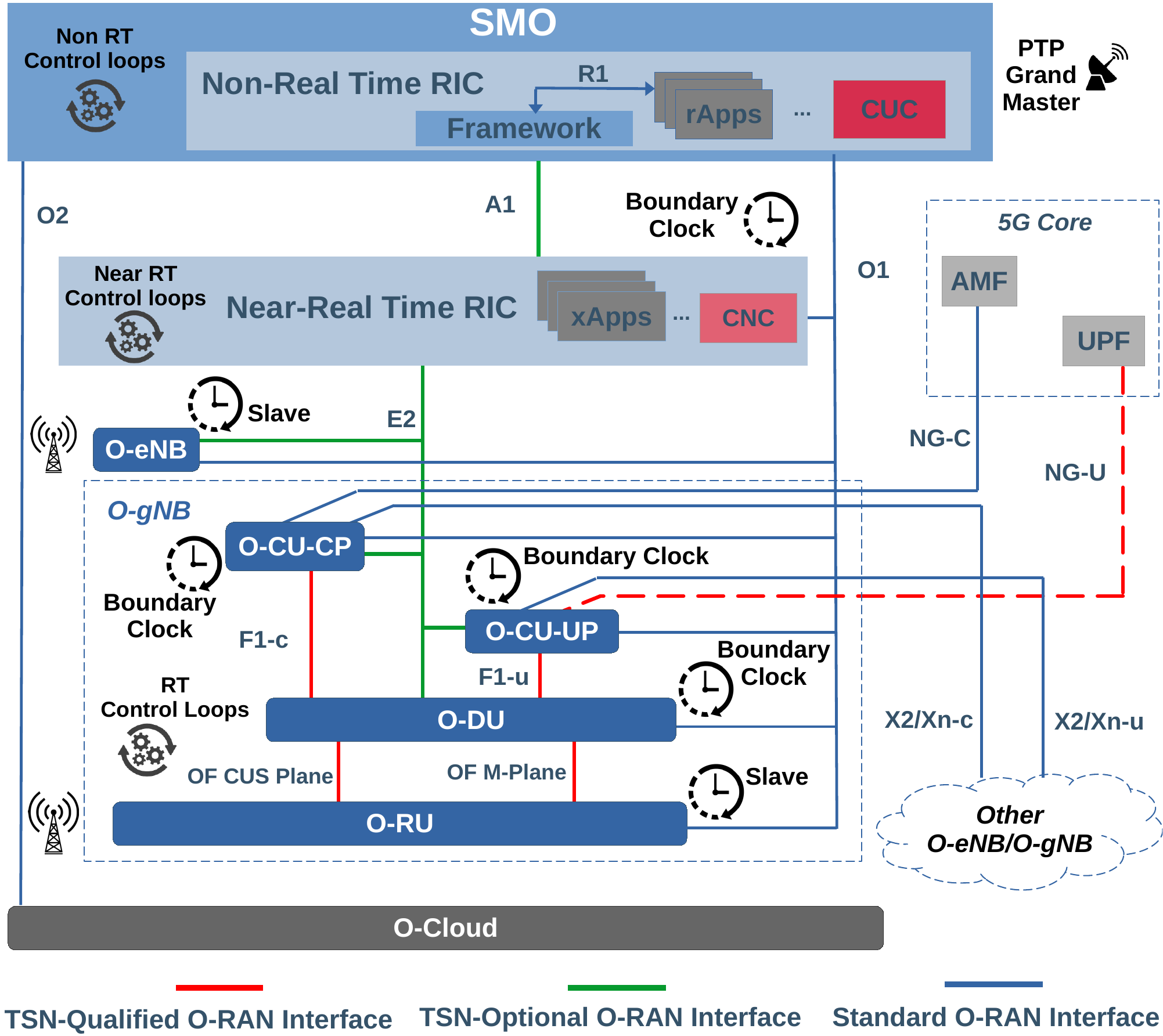}
        \caption{O-RAN architecture and its integration with TSN.}
        \label{fig:arch}
\end{figure}

The O-RAN time-critical interfaces (marked in red and named after \textit{TSN-Qualified} interfaces in Fig.~\ref{fig:arch}) are the ones that can highly benefit from TSN. This mainly includes the OpenFronthaul Control, User and Synchronization (CUS) plane and Management (M) plane, the F1-c/F1-u interfaces that interconnect O-DUs with O-CUs and the transport data plane interface (NG-U) that interconnects O-RAN with the core network.
The NG-U interface is by default time-sensitive compliant since it is usually implemented through large-scale optical transport networks that provide low-latency communication and determinism through wavelength isolation. 
However the OpenFronthaul CUS and M planes and F1-u/F1-c interfaces, which are under the umbrella of the near-RT RIC, often use overdimensioned point-to-point links to ensure the latency budget. To reduce costs, operators can adopt TSN to deploy these interfaces over an integrated Ethernet transport network that would ensure low-latency with truly deterministic guarantees for time-critical flows. Even when sharing the network with other time-critical traffic flows, such as neighboring fronthaul payload, critical edge services or external Best-effort (BE) traffic. 

On the other hand, implementing TSN requires additional mechanisms to control the time-sensitive streams. IEEE 802.1Qcc~\cite{IEEE8021Qcc} (amendment where stream management enhancements are introduced) defines three models for TSN user/network configuration: a) \textit{fully-distributed}, b) \textit{hybrid}, and c) \textit{fully-centralized}. While the \textit{fully-distributed} model could have some niche applications (e.g., marketplace-based decentralized green radio use cases), the \textit{hybrid} and \textit{fully-centralized} models are the ones that may fit better in the O-RAN architecture. Both TSN models envisage the use of a Centralized Network Configuration (CNC) entity, which has full view of the physical topology and can compute schedules and paths for the time-sensitive flows. The CNC acts as a Software Defined Networking (SDN) controller with extended TSN functionalities. The O-RAN architecture may natively support a CNC through its deployment in the near-RT RIC as an \textit{xApp}, running a near-RT control-loop that continuously audits the performance of the flows, and re-configuring the network when the requirements are not met. 

While the end-nodes (i.e., talkers/listeners in TSN nomenclature) directly express their flow requirements to the CNC in the \textit{hybrid} model, all flow requirements are expressed and managed by the Centralized User Configuration (CUC) entity in the \textit{fully-centralized} model. The role of the CUC is to discover end-nodes, retrieve their flow requirements, and configure their TSN features. Since such operations are usually not frequent (e.g., commissioning of new equipment), the CUC may be implemented in the non-RT RIC as an \textit{rApp} within the O-RAN architecture. 
Figure~\ref{fig:arch} includes possible locations for the CNC, CUC, and Boundary/Master-clocks, which are required to synchronize the TSN-enabled elements. 

Additionally, the A1 and E1 interfaces are not time-critical because of their time scales, however they can also benefit from bounded latency through shared Ethernet networks (marked in green and named after \textit{TSN-Optional} interfaces in Fig.~\ref{fig:arch}). Although the timing requirements are not as strict as the TSN-Qualified interfaces, interference-less and timely control would enable shorter and more accurate control-loops.

\section{Ecosystem and deployment options}\label{sec:deployment}

\subsection{The time-sensitive O-RAN ecosystem}
\subsubsection{OpenFronthaul}

Traditional LTE and NR RAN deployments relied on the CPRI/eCPRI standards for the fronthaul interfaces~\cite{eCPRIv2requirements}. Although such interfaces are standardized, the extensive modifications done by individual vendors cause a de-facto lock-in between their RUs and DUs. To ensure openness, O-RAN defined the OpenFronthaul interface to ensure multi-vendor coexistence between O-DUs and O-RUs. The O-RAN OpenFronthaul is an entirely packet-based interface which can leverage Ethernet to transport fronthaul payloads, although it can also use Wavelength Division Multiplexing (WDM) optical networks. 
OpenFronthaul conforms to the eCPRI framework and shares its latency models and, by leveraging TSN, it may transport time-critical payloads along with other time-critical/BE traffic without performance degradation in a cost-effective manner over Ethernet. OpenFronthaul is based on two new interfaces, the CUS-plane and M-plane.

The OpenFronthaul CUS-Plane has the highest priority in the OpenFronthaul, that is, non-preemptable Expedited Forwarding (EF).
 It consists of i) the \textit{U-Plane}, which carries the user plane IQ data precoded in the frequency domain, ii) the \textit{C-Plane}, which  controls the data transfer, manages the scheduling, numerology and Physical Random Access Channel (PRACH), and processes beam-forming commands, and iii) the \textit{S-Plane}, which is responsible for the time, frequency and phase synchronization between the O-DUs and O-RUs. O-RUs and O-DUs need to be tightly synchronized in order to be able to perform latency critical operations such as Time Division Duplexing (TDD), Carrier Aggregation (CA), MIMO or handovers. TSN mechanisms that require synchronization such as 802.1Qbv may leverage the S-Plane as well.
 
\begin{figure}
        \centering
        \includegraphics[width=0.49\textwidth]{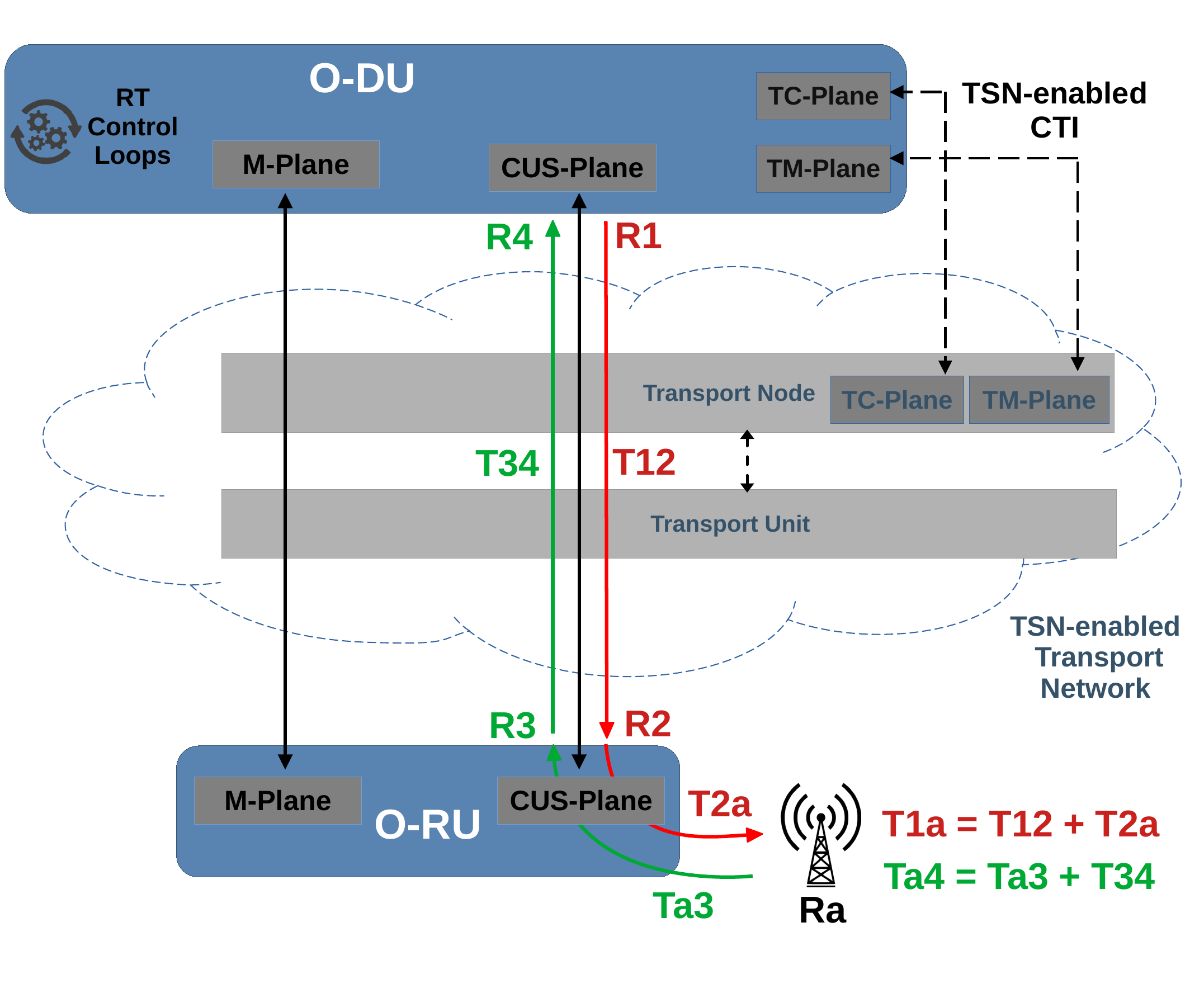}
        \caption{OpenFronthaul interface planes and reference points for delay management.}
        \label{fig:of}
\end{figure}

Conversely, the \textit{M-Plane} is used for providing Fault, Configuration, Accounting, Performance and Security (FCAPS) support. Some of its functions are life-cycle management, software updates or fault monitoring, and therefore it does not have the latency requirements of the CUS-Plane. Instead, M-Plane has preemptable Assured Forwarding (AF) priority. A summary of the OpenFronthaul logical planes is depicted in Fig.~\ref{fig:of}.

Additionally, O-RAN extends its OpenFronthaul interface with the Cooperative Transport Interface (CTI). While TSN is able to operate a fabric of Ethernet switches with reservation-based features, CTI supports cooperation with external, shared resource-allocation-based networks (e.g., PON, WDM or DOCSIS). In this sense, CTI enables the identification and classification of fronthaul flows by the Transport Nodes (TNs) and Transport Units (TU), and allows them to process each flow according to their corresponding timing requirements. 
A TSN-enabled O-DU may interact with the CTI Transport Control (TC) and Transport Management (TM) planes through a TSN Agent implemented as a~\textit{dApp}, a concept introduced in~\cite{d2022dapps} to deploy applications in the O-gNBs.

\subsubsection{The O-RAN Midhaul: F1-u and F1-c}

F1-u/c interfaces are commonly considered as the midhaul and are defined by the 3GPP TS 38.470 standard. 
O-RAN may also consider 3GPP F1-u and 3GPP F1-c as time sensitive interfaces as they are under the umbrella of the near-RT RIC. F1-u/c have less stringent latency and PDV requirements and thus can be transported over a number of IP-enabled transport technologies such MPLS or SRv6.

\subsection{Requirements and Deployment Options}

The TSN-enabled deployment scenarios in O-RAN are heavily influenced by the timing requirements imposed by the time-sensitive interfaces. The two main requirements that must be fulfilled are the one-way delay measurement, namely, latency, and the time synchronization error, namely, Time Alignment Error (TAE). Different deployment options are allowed as long as the delay and the TAE are bounded on a maximum value. Since the tightest requirements are those of the fronthaul (e.g., typical end-to-end one-way latency is 100$\mu$s for the U-plane, and 1.5$\sim$10ms for F1-u), next we focus on the OpenFronthaul.

The timing in the OpenFronthaul interface follows the eCPRI delay management and it is tightly coupled with the air interface. Figure~\ref{fig:of} shows the location of the standard eCPRI reference points (e.g., R1, R2, T1a, T2a, etc.), used to calculate the delays. The main idea is to ensure that TX/RX windows of the O-DU are properly aligned to support the O-RU transport characteristics, mainly derived from transport constraints and 5G NR options, for example, Channel Bandwidth and Subcarrier Spacing (SCS), U-Plane/C-Plane timing relations, and retransmission timing of the Hybrid Automatic Repeat reQuest (HARQ). 

\begin{figure}
        \centering
        \includegraphics[width=0.50\textwidth]{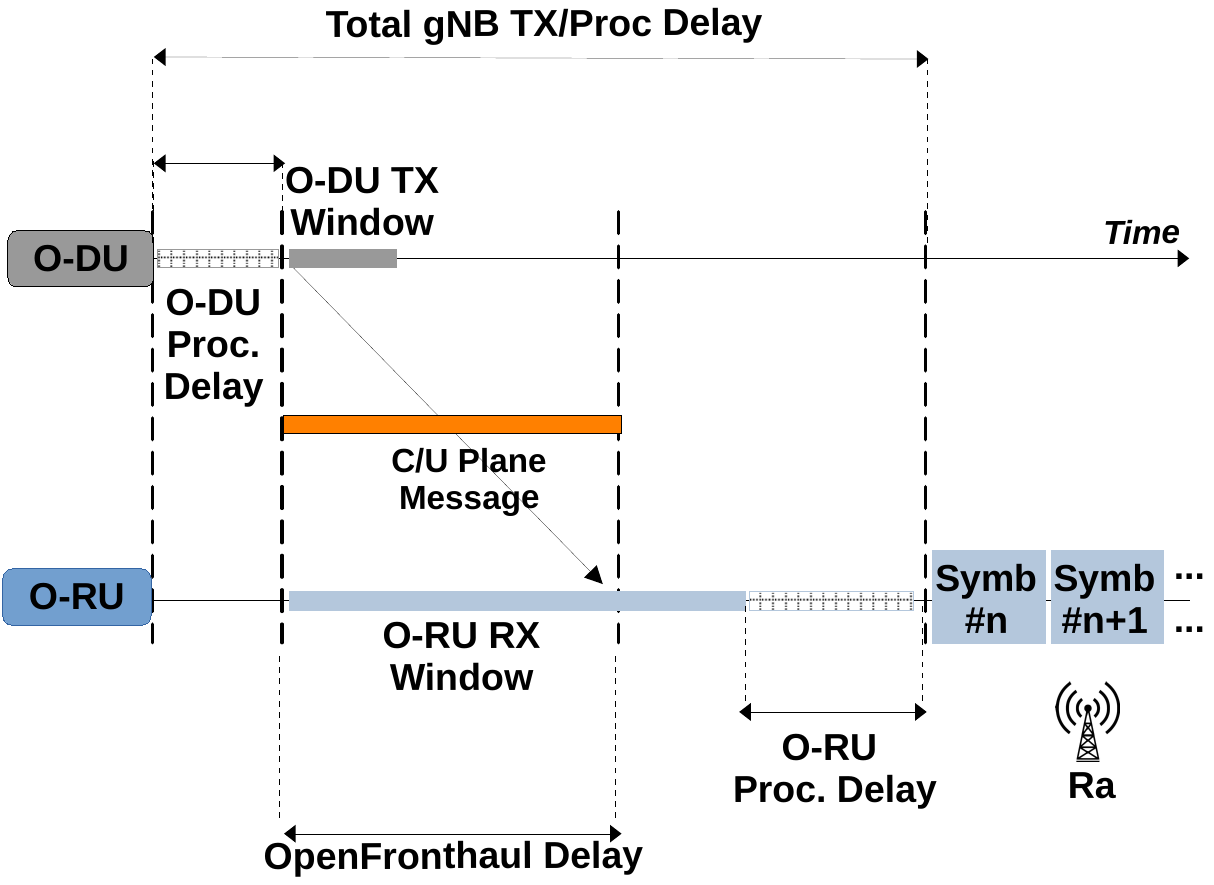}
        \caption{Timing relations per U/C message in DL direction.}
        \label{fig:delaymgnt}
\end{figure}

An example of delay management is presented in Fig.~\ref{fig:delaymgnt}. The $TX_{Window}$ and $RX_{Window}$ are set accounting for the O-DU processing delay, medium access delay, OpenFronthaul delay (switching delay, propagation delay, etc.) and O-RU processing delay (and their estimated maximum fluctuations). The $TX_{Window}$ is defined as $T1a_{max} - T1a_{min}$, where $T1a$ is the elapsing time from a bit leaving the O-DU ($R1$) until is received at the antenna interface ($Ra$). The \textit{max} and \textit{min} values account for the fluctuation delays, and is typically predefined based on the equipment capabilities.
Equally, the $RX_{Window}$ is defined as $T2a_{max} - T2a_{min}$. In this sense, the $RX_{Window}$ must be at least long enough so that it can cope with the worst case within the $TX_{Window}$ (i.e., $RX_{Window}$~$\geq$~$TX_{Window}$ + {OpenFronthaul maximum delay}).

To support such stringent time requirements through TSN, OpenFronthaul may leverage 802.1CM~\cite{IEEE8021CM2018}. 802.1CM defines mechanisms for transporting time-sensitive fronthaul streams over bridged networks\footnote{Fronthaul networks other than bridged networks are outside the scope of 802.1CM.}. It consists of two interface classes:

\begin{itemize}
\item[•] \textbf{Class 1:} it refers to fronthaul interfaces that use CPRI to periodically send IQ data, regardless there is actually traffic from the O-RU or not. This results in a Constant Bit Rate (CBR) stream sent over periodic time-windows. 
This class may highly benefit from 802.Qbv scheduling to provide deterministic CPRI support over Ethernet networks.  
\item[•] \textbf{Class 2:} it refers to fronthaul interfaces that use eCPRI. The IQ data is transported in a more traffic-efficient packet-based stream. Unlike Class 1, data is only exchanged when there is actual user data from the O-RU.
\end{itemize}

Additionally, 802.1CM requires disabling Low-Power Idle (LPI) mode to avoid the delay (wake time) when the LPI is de-asserted, and forbids the use of control protocols such as MAC control Pause (802.1Qbb). 
On the other hand, 802.1CM clashes with other priority-based flow control mechanisms defined in 802.1Q operating on the priorities required by fronthaul traffic. However, for those scenarios, applying 802.1Qbv traffic scheduling would allow for the co-existence of several time-sensitive flows with the highest priority. Finally, transport network bridges must meet some requirements (e.g., a minimum of 1 Gb/s ports, 2000 byte maximum PDU size, full-duplex point-to-point links, etc.), and are able to transport non-fronthaul traffic as long as the fronthaul requirements are met.

802.1CM also defines two fronthaul profiles to fulfill the targets of the above-mentioned classes. Profile A operates through strict priority queuing, there is not bandwidth limitation and the maximum frame size is 2000 bytes. Profile B extends Profile A with 802.1Qbu Frame Preemption to reduce interference of non-fronthaul traffic over fronthaul traffic. The worst-case preemption latency is defined by a minimum fragment size of 124 bytes (1080ns in a 1Gb/s link).

Since synchronization between O-RUs is crucial, O-RAN specifies four types of OpenFronthaul Low-Layer Split (LLS) topology configurations to provide synchronization~\cite{oranSyncArchWP9}: 

\begin{itemize}
\item[•] LLS-C1: This topology configuration assumes a direct point-to-point connection between O-DU and O-RU. The O-DU acts as Telecom Grand-Master (T-GM) and synchronizes through SyncE and PTP the O-RU, which acts as Telecom Slave-Clock (T-TSC).
\item[•] LLS-C2: There is a multi-hop connection between the O-DUs and the O-RUs through a network of Ethernet switches, which act as Telecom Boundary-Clocks (T-BCs). The maximum number of hops is limited by the error contributions of all clocks in the chain (i.e., including TAE and frequency error in the air interface). 
\item[•]LLS-C3: It supports network timing distribution from one or more external T-GMs located in the transport network. O-DUs (optionally) and O-RUs act as T-TSCs.
\item[•]LLS-C4: It assumes all O-DUs and O-RUs have their own GNSS-based clocks and thus, the synchronization is not done through the transport network.
\end{itemize}

Additionally, O-RAN supports several timing profiles to ensure synchronization interoperability among O-RAN network elements. They are: Full-path Timing Support (FTS) profile (ITU-T G.8275.1), where all elements between the T-GM and the T-TSC are PTP aware devices, Partial Timing Support (PTS) profile (ITU-T G.8275.2), where some of the elements are not PTP aware, and Assisted Partial Timing Support (APTS) profile (ITU-T G.8275.2), where elements use GNSS-based clocks as primary reference source and only use PTP as backup timing source. FTS and PTS are normally used for LLS-C1/C2/C3 while APTS is intended to be used in LLS-C4 topologies. For applications such as fronthauling of Category B O-RUs over TSN networks with high TAE requirements (i.e., $<$ 260ns), O-RAN recommends the use of the FTS profile~\cite{oranSyncArchWP9}.

\begin{figure*}
\centering
\begin{subfigure}{.5\textwidth}
  \centering
  \includegraphics[width=1\linewidth]{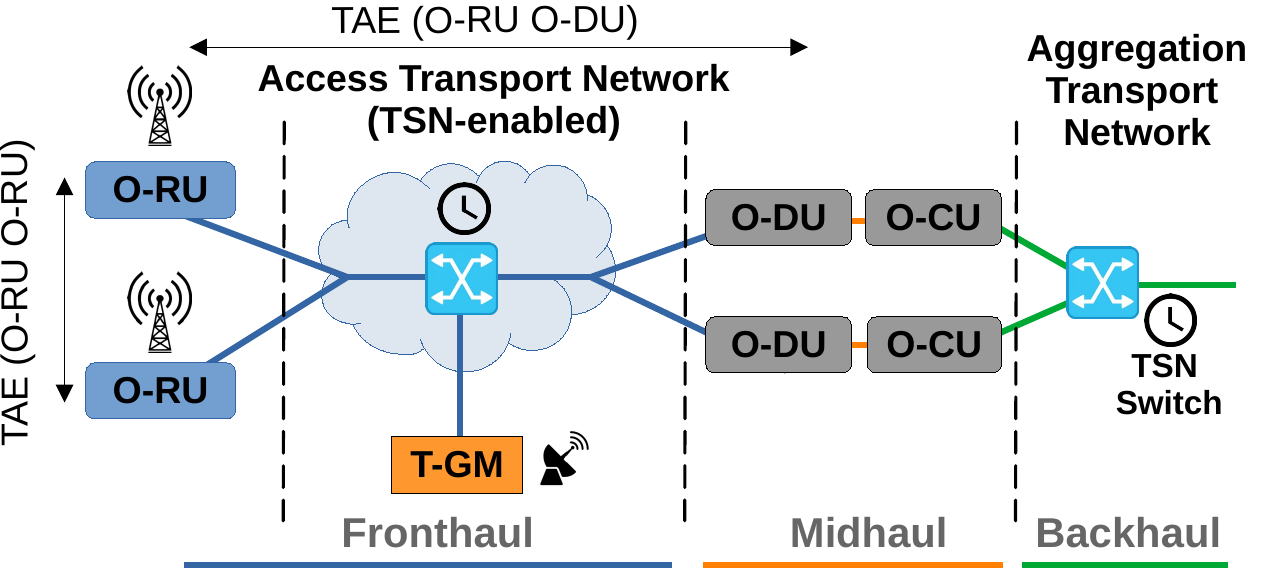}
  \caption{}
  \label{fig:sc1}
\end{subfigure}%
\begin{subfigure}{.5\textwidth}
  \centering
  \includegraphics[width=1\linewidth]{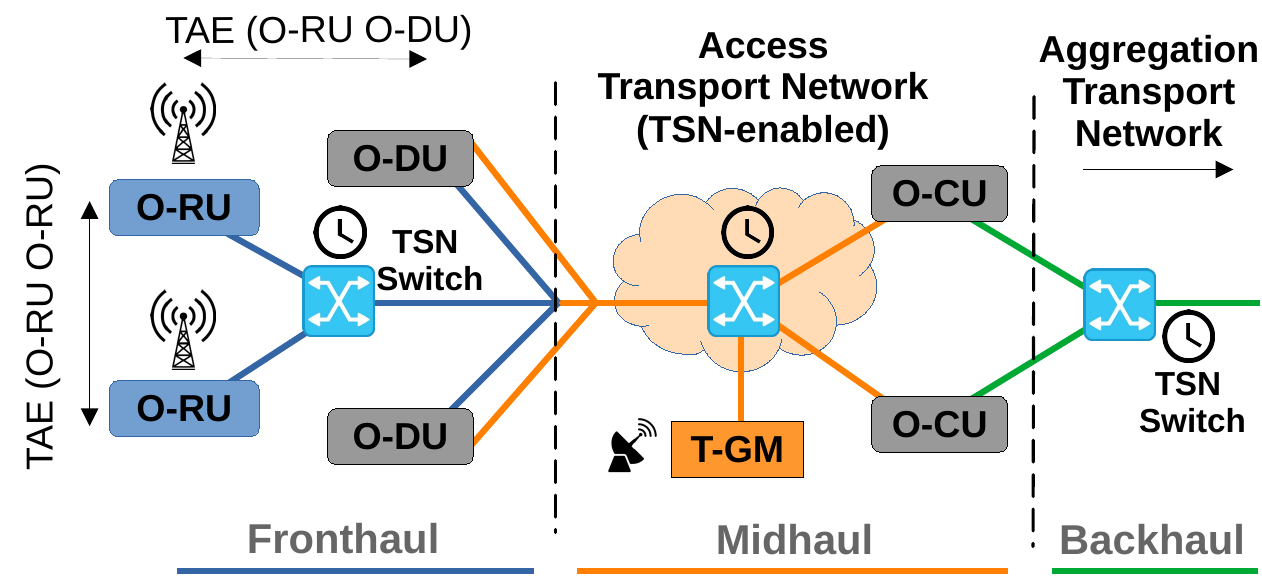}
 \caption{}
  \label{fig:sc2}
\end{subfigure}
\caption{Different deployment options and their effect on the TAE according to the O-RU and O-DU locations.}
\label{fig:synctopos}
\end{figure*}

Choosing the right timing profile, LLS topology configuration and location of the clocks in a O-RAN deployment is non-trivial. First, strict TAE requirements limit link distances and forbid certain topologies. Secondly, while centralized approaches are simpler and have lower costs since only few GMs are needed, decentralized ones do not need synchronization from the network. However they involve higher costs and are prone to GNSS signal jamming or spoofing.
Two design choices for a typical O-RAN deployment (i.e., Scenario 1 with co-located O-DU/O-CU, and Scenario 2 with co-located O-RU/O-DU, both configured as LLS-C3~\cite{oranOFCUS}) are given in Fig.~\ref{fig:synctopos}.
Depending on the deployment scenario, the O-RU/O-RU TAE and the O-RU/O-DU TAE will vary significantly.

\subsection{O-RAN Software Community and integration with TSN}
The O-RAN Software Community (O-RAN SC) is a partnership between the O-RAN Alliance and the Linux Foundation whose main goal is to support the software development of an Open RAN solution available to everyone. A proper alignment between the O-RAN community-driven software developments projects and the TSN working groups is therefore essential to achieve a successful integration. 
The O-RAN SC is mainly composed by the following software projects: 
\begin{itemize}
    \item RIC Applications (RICAPP): Includes open source sample \textit{xApps} and platform applications for integration, testing and demonstrations.
    \item Near-RT RIC (RIC): Includes the near-RT RIC implementation using OpenAPI to support \textit{xApps}.
    \item Non-RT RIC (NONRTRIC): Includes the non-RT RIC implementation focusing on its interoperability through the A1 interface and closed-loop use cases.
    \item Operations and Maintenance (OAM): Provides reference sample implementations according to the operations and maintenance O-RAN documents.
    \item O-RAN Central Unit (OCU): Provides a baseline implementation for the O-CU, focusing on a basic E2 interface to enable integration and testing between RIC and OCU.
    \item O-RAN Distributed Unit High Layers (ODUHIGH): Implements the functional blocks of the 5G NR L2 protocol stack defined for the standalone (SA) mode. 
    \item O-RAN Distributed Unit Low Layers (ODULOW): Provides an initial implementation of the functional blocks of the 5G NR L1 protocol stack.
    \item Infrastructure (INF): Provides initial infrastructure building blocks to properly deploy O-RAN network functions components (i.e., real-time platform deployment).
    \item Service Management and Orchestration (SMO): 
    It implements the O1 (RAN Network Functions configuration and management) and the O1/VES (RAN-generated events gathering) interfaces.
\end{itemize}

We focus on those containing time-sensitive communication interfaces. 

\subsubsection{O-RAN Fronthaul Interface Library}
The O-RAN Fronthaul Interface (FHI) library is part of the ODULOW software project. This library, built on top of DPDK\footnote{\url{https://www.dpdk.org/}}, performs U-plane and C-plane functionality according to the OpenFronthaul specification. 
It allows ODULOW software components to use the C-Plane and U-Plane of the O-RU, and is expected to communicate TTI events, symbol time, C-plane information as well as IQ Samples data. The current implementation relies on the xRAN library\footnote{\url{https://docs.o-ran-sc.org/projects/o-ran-sc-o-du-phy/en/latest/xRAN-Library-Design_fh.html}} to provide support for transporting IQ samples between O-DU and O-RU. The library defines packet formats to transport radio samples according to the O-RAN specification. The FHI library can be made agnostic to the network interfaces, leaving TSN-enabled tasks such as frame preemption or traffic scheduling to the kernel. However, for some ultra-low latency use cases, it may be required to move such tasks to DPDK to ensure faster processing times.

For time synchronization, the xRAN library currently supports configurations LLS-C1 and LLS-C3, and primary/secondary PTP configurations are expected to satisfy only O-DU requirements, while providing best-effort primary PTP for O-RU. This may not be sufficient to accomplish the S-plane requirements, which could lead to require additional Boundary-Clocks. Finally, while O-RAN fronthaul data can be transported over Ethernet or IPv4/IPv6, the current xRAN library implementation only supports Ethernet through VLANs.

\subsubsection{O-DU High Layers}
The O-DU high layers software project includes the required modules to implement the functional blocks of the 5G NR L2 in different software entities:
\begin{itemize}
    \item DU-APP: configures and manages all the operations of the O-DU by interacting with the OAM, over the O1 interface for configuration, alarms and performance management, the O-CU, on the F1 interface over SCTP, and the RIC, on the E2 interface over SCTP.
    \item 5G-NR-RLC: provides services for communication with the O-CU through the DU-APP.
    \item 5G-NR-MAC: responsible for sending/receiving data on the various logical channels, scheduling grants, multiplexing/de-multiplexing, and Femtocell Application Programming Interface (FAPI) communications.
    \item O1-Module: main agent implementing the core functionality of the O-DU High Layers software.
\end{itemize}

For the integration of the standard centralized/hybrid TSN architecture on the O-DU High Layers project, an additional south-bound interface between the O-DU and CNC is required. It could be O1, or an extension of E2, and its aim would be to provide the O-DU with scheduling information that will be used for i) adjust its own TX/RX windows facing the O-RU, ii) align such windows with the schedules of the Ethernet switches involved in the fronthaul path.

\begin{table*}[t]
  \centering
  \footnotesize
  \caption{Mapping of available O-RAN interfaces with their requirements and technologies.}
  \begin{tabular}{c|cccccccccc}
    
    { }      & \makecell{Max.\\ Delay}  & \makecell{Max. \\FLR}  & Encapsulation & {Ethernet} &  \makecell{PON\\WDM} &  {DOCSIS}  & {Microwave} & {mmWave} & \makecell{TSN\\Qualified} & \makecell{TSN\\Optional} \\
    \hline
    OF C  & 1 ms  & $10^{-7}$  & VLAN/eCPRI &Yes & Yes & No  & No & Yes & \checkmark &   \\
    OF U  & 25 $\mu$s - 1 ms  & $10^{-7}$  & VLAN/eCPRI  & Yes & Yes & No & No & Yes& \checkmark &   \\
    OF S  & 25 $\mu$s - 500 $\mu$s  & $10^{-7}$  & VLAN/PTP  & Yes & Yes & No & No & Yes& \checkmark &  \\
    OF M  & 100 ms  & $10^{-6}$ & VLAN/NETCONF & Yes & Yes & Yes & Yes & Yes&  & \checkmark \\
    F1-c  & 1.5-10 ms  & N/A  & VLAN/F1AP  & Yes & Yes & Yes (LLX) & Yes & Yes& \checkmark &  \\
    F1-u  & 1.5-10 ms  & N/A & VLAN/GTP-U & Yes & Yes & Yes (LLX)  & Yes& Yes& \checkmark &  \\
    E2  & 10 ms  & N/A  & VLAN/E2AP & Yes & Yes & Yes (LLX) & Yes & Yes &  & \checkmark \\
    A1  & 500 ms  & N/A  & VLAN/A1AP & Yes & Yes & Yes & Yes & Yes &  & \checkmark \\
    NG-U  & 1-50ms  & N/A  & VLAN/GTP-U & Yes & Yes & Yes & Yes & Yes &  & \checkmark \\
  \end{tabular}
  \label{tab:summary}
\end{table*}

\section{Discussion}\label{sec:discussion}
To date, and inheriting from traditional 3GPP RANs, proposed O-RAN approaches are rather conservative and often consider RAN transport networks as overdimensioned point-to-point or static point-to-multi-point links to deal with the strict latency requirements in the time-sensitive interfaces.
Aiming to improve cost-efficiency, future O-RAN network-sharing use cases in Ethernet transport networks where multiple O-RUs/O-DUs/O-CUs use the same network to communicate one each other, require however TSN mechanisms to ensure packet delivery with fair and deterministic performance. As a summary, Table~\ref{tab:summary} provides an overview of the O-RAN interfaces currently available in the O-RAN SC that may leverage TSN, including its encapsulation modes, latency requirements and transport technology options.

The most restrictive interfaces are the OpenFronthaul CUS-Plane. While the U-Plane and S-Plane may require up to 25$\mu$s for ultra-low-latency performance use cases, typical scenarios usually require 100$\mu$s. Thus, this excludes the use of DOCSIS for its use in the CUS-plane, even when using Low-Latency Xhaul (LLX), and Microwave, where typically only bands above 60GHz achieve latencies lower than 250-500$\mu$s. Conversely, OpenFronthaul M-Plane has more relaxed requirements. It is worth mentioning that O-RAN also considers non-ideal fronthaul networks for low UE density and slow-fading scenarios (e.g., femto-cells in rural areas) where latencies in the order of few milliseconds may have a negligible impact in the performance~\cite{oranOFCUS}. In the F1-u/F1-c, NG-U, A1 and E2 interfaces the latency requirements are less strict, therefore more transport technologies options are available. 

Additionally, although typical LTE and 5G NR deployments are nowadays based on WDM, some operators may choose Ethernet to save costs. This is demonstrated in~\cite{gonzalez2019integrating}, where CPRI fronthaul traffic safely coexists with backhaul traffic in the same wired switching network as a way to reduce deployment costs. In fact, the increased number of sectors in 5G NR results in many channels to be transported in the fronthaul along with the inefficient 4G CPRI legacy channels. Such a large number of channels presents a challenge in terms of costs, since traditional approaches based on WDM usually require many expensive 10$\sim$25Gb/s optics. However, by using TSN over Ethernet in a single fiber, channels can be multiplexed and CPRI, eCPRI and external traffic can coexist achieving a reduction of Total Cost of Ownership (TCO) of 30\%\footnote{A detailed economic analysis of a shared fronthaul deployment can be found here: \url{https://ec.europa.eu/research/participants/documents/downloadPublic?documentIds=080166e5b77d8419&appId=PPGMS}}. Last and also relevant, sometimes operators just need to support legacy low-speed Ethernet-based RAN transport networks (1GE or even 100BASE-T). In such scenarios, enabling TSN would also help to manage the strict latency budgets.

Unlike the midhaul and backhaul interfaces, the OpenFronthaul is the only interface that sets requirements in terms of Frame Loss Ratio (FLR). This is because, while the loss of a frame in the U-Plane will generally impact only a specific symbol, the loss of a C-Plane frame could possibly impact an entire slot's worth of data. Therefore, FLR over the considered transport options is often considered negligible (except in Microwave links). Finally, the time-critical interfaces shown in Table~\ref{tab:summary} may all benefit from TSN whenever the network is shared with external traffic, but it is in the OpenFronthaul CUS-Plane and F1-u/F1-c interfaces where TSN could play a major role. 

\subsection{802.1Qbu and 802.1Qbv}

Applying 802.1Qbu Frame Preemption is straightforward in O-RAN. It is specified in the 802.1CM Profile B and it has great impact in reducing PDV on sub-10GE links. However, standards may have neglected the benefits of applying 802.1Qbv's scheduling approach to share cost-effective fronthaul networks with theoretically zero PDV. This is in part due to the risk of resource wasting if 802.1Qbv slots are not used. Yet this can be mitigated by jointly applying 802.1Qbu. We identify the joint 802.1Qbv/802.1Qbu implementation as a lower-cost alternative to provide time division multiplexing over Ethernet, as other solutions such as FlexE and ITU-T G.8312 do. However, while the latter lacks the statistical multiplexing gain of pure strict priority approaches, 802.1Qbv in combination with 802.1Qbu can marge both worlds, ensuring deterministic delay and low delay variation, and allowing non-TSN preemptable traffic to use the excess capacity at the same time. This mechanism would enable, for example, to use the same Ethernet network for both 4G CPRI and 5G eCPRI fronthaul traffic in a cost-effective, flexible and scalable manner. Additionally, there is a potential for aligning 802.1Qbv cycles to TTIs to reduce queuing time, and in optimizing ON-OFF traffic of 4G CPRI-legacy deployments by bounding OFF times within 802.1Qbv cycles. Finally, 802.1Qbv could also be leveraged to adapt network latency budgets according to computational capabilities in the O-DU (e.g., dynamically prioritizing frames that have been delayed because of the virtualization overhead), and to possibly enable dynamic functional splits.

\subsection{To Schedule or not to Schedule}

However, the debate is still open. Despite the benefits that scheduling-based TSN techniques may bring to O-RAN time-critical interfaces, there is also an inherent increase in network management complexity. First, the CUC and CNC need to be deployed in the O-RAN architecture and their interactions need to be defined. The most logical options are to deploy the CUC as an \textit{rApp} in the SMO or non-RT RIC and the CNC as a \textit{xApp} in the near-RT RIC. However, the 802.1Qcc specification leaves the CUC-CNC interaction in the vendors' hands, which may require extra standardization effort from the O-RAN Alliance side. Also, while 802.Qbu only involves the re-assembling of the preempted frames in the other end of the link where it is applied, applying 802.1Qbv increases implementation complexity in the switches, and propagating schedule updates involves extra control overhead in the O1 or E2 interface. Secondly, computing schedules at the CNC entails heavy CPU-intensive loads, and near-real-time performance may be challenging for exact algorithms when the network fluctuates significantly~\cite{steiner2018traffic}. However, this also opens interesting research paths for Artificial Intelligence and Machine Learning (AI/ML) algorithms to support CNC scheduling tasks. This includes research on both the intelligent scheduling algorithm itself and the intelligent decision algorithm that decides which algorithm suits best and where it can be executed most efficiently at any given moment for a particular scenario. 

\subsection{Additional TSN standards}

Besides 802.1Qbu and 802.1Qbv, O-RAN time-sensitive interfaces may also leverage other mechanisms from the TSN toolbox. For instance, IEEE 802.1Qci Per-Stream Filtering and Policing provides improvements for isolation and coordination of traffic flows, and IEEE 802.1Qch Cyclic Queuing and Forwarding, binds bounded latency to the hop-count and cycle time. Both 802.1Qci and 802.1Qch can be jointly used with 802.1Qbv to achieve more accurate control over the schedules. However, other amendments such as IEEE 802.1CB Frame Replication and Elimination for Reliability or IEEE 802.1Qca Path Control and Reservation may only be relevant on RAN transport networks with high frame error rates (e.g., based on Microwave links) or with complex heterogeneous topologies.

\section{Conclusion}

O-RAN time-critical interfaces certainly benefit from applying TSN to provide determinism to latency-sensitive and latency variation-sensitive streams over cost-effective Ethernet transport networks. Therefore, reduced operator costs in terms of increased deployment efficiency and less infrastructure can be achieved by replacing overdimensioned dedicated links with general-purpose TSN-enhanced Ethernet.  
This work summarizes the key aspects of the O-RAN/TSN blend, while providing an overview of the joint requirements and deployment options. Finally, we discuss about the potential opportunities and challenges of bringing O-RAN and TSN closer together for the purpose of allowing the O-RAN Alliance to fully reach the openness and virtualization levels it aims to.

\section{Acknowledgment}

This work has been supported by the European Commission through Grant No. 101017109 (DAEMON project) and Grant No. 101097083 (BeGREEN project), by the Spanish Ministry of Economic Affairs and Digital Transformation and the European Union – NextGeneration EU, in the framework of the Recovery Plan, Transformation and Resilience (PRTR) (Call UNICO I+D 5G 2021, ref. number TSI-063000-2021-3-Open6G), and by the CERCA Programme from the Generalitat de Catalunya.

\bibliographystyle{IEEEtran} 
\bibliography{orantsn}

\begin{IEEEbiographynophoto}{Esteban Municio}
received his Ph.D. degree from the University of Antwerp at imec (Belgium) in 2020. He then joined imec as postdoctoral researcher for two years. Since January 2022, he has been with i2CAT, where currently he is a research scientist in the AI-driven Systems group. His research interests are in the field of programmable open networks, Time-Sensitive Networking and ultra-reliable Industrial IoT.
\end{IEEEbiographynophoto}

\begin{IEEEbiographynophoto}{Gines Garcia-Aviles}
received his Ph.D. degree from the University Carlos III of Madrid. Since January 2021, he has been with i2CAT, where currently he is a research scientist in the AI-driven Systems group. Gines’ research interests are RAN Softwarization, SDN and NFV technologies for future cellular networks.
\end{IEEEbiographynophoto}

\begin{IEEEbiographynophoto}{Andres Garcia-Saavedra}
received his Ph.D. degree from the University Carlos III of Madrid in 2013. He then joined Trinity College Dublin, Ireland, as a research fellow. Since July 2015, he has been with NEC Laboratories Europe, where currently he is a principal research scientist. His research interests lie in the application of fundamental mathematics to real-life wireless communication systems.
\end{IEEEbiographynophoto}

\begin{IEEEbiographynophoto}{Xavier Costa-Pérez}
is an ICREA Research Professor, Scientific Director at the i2CAT Research Center, and Head of 5G/6G Networks R\&D at NEC Laboratories Europe. He has served on the Organizing Committees of several conferences, published papers of high impact and holds tenths of granted patents. He received his Ph.D. degree in telecommunications from the Polytechnic University of Catalonia, Barcelona, and was the recipient of a national award for his Ph.D. thesis.
\end{IEEEbiographynophoto}

\end{document}